\documentclass{jpp}
\usepackage{graphicx}
\usepackage{tensor}
\usepackage[markup=draft]{changes}
\definechangesauthor[name=JCP, color=red]{JCP}
\usepackage{bm}
\usepackage{academicons,xcolor}
\usepackage[colorlinks = true,
            linkcolor = blue,
            urlcolor  = blue,
            citecolor = blue,
            anchorcolor = blue]{hyperref}

\usepackage[utf8]{inputenc}
\usepackage[T1]{fontenc}
\usepackage{amsmath,color}

\renewcommand{\vec}{\bm}

\newcommand{\sun}{\odot}
\newcommand{\diff}[2]{\frac{\mathrm{d}#1}{\mathrm{d}#2}}	
\renewcommand{\tt}{\tilde{\theta}}							
\newcommand{\dotline}[1]{\makebox[#1]{\dotfill}}

\newcommand{\rms}{\rm rms}
\renewcommand{\d}{\textrm{d}}								

\renewcommand{\div}{\nabla\bcdot}							

\renewcommand{\vec}[1]{\mathbf{#1}}

\renewcommand{\div}[1]{\vec\nabla\cdot#1}

\newcommand{\eb}{\vec{\hat b}}

\newcommand{\aave}[1]{\left\langle #1\right\rangle}

\newcommand{\orcid}[1]{\href{https://orcid.org/#1}{\textcolor[HTML]{A6CE39}{\aiOrcid}}}

\newcommand{\apj}{Astrophys.~J.}
\newcommand{\jgr}{J. Geophys. Res.}
\newcommand{\apjs}{Astrophys. J. Suppl.}
\newcommand{\apjl}{Astrophys. J. Lett.}
\newcommand{\solphys}{Solar Phys.}

\def\solphys{Sol. Phys.}
\def\ssr{Sp. Sci. Rev.}

\def\apj{Astrophys. J.}
\def\apjl{Astrophys. J. Lett.}
\def\aap{Astron. Astrophys.}

\def\apjs{Astrophys. J. Suppl.}
\def\jgr{J. Geophys. Res.}

\title{How Alfv\'en waves energize the solar wind: heat vs work}

\author{Jean C.\ Perez\aff{1}\corresp{jcperez@fit.edu}, Benjamin D.\ G.\ Chandran\aff{2}, Kristopher G. Klein\aff{3},
  and Mihailo M. Martinovi\'{c}\aff{3,4}}

\affiliation{\aff{1}{Department of Aerospace, Physics and Space
    Sciences, Florida Institute of Technology, Melbourne, Florida, USA}\aff{2}{Department of Physics and Astronomy, University of New
    Hampshire, Durham, New Hampshire 03824,  USA },\aff{3}{Lunar and
    Planetary Laboratory, University of Arizona, Tucson, AZ 85721,
    USA},\aff{4}{Laboratoire d’Etudes Spatiales et d’Instrumentation
    en Astrophysique, Observatoire de Paris, Meudon, France}}

\begin{document}

\maketitle

\begin{abstract}
  A growing body of evidence suggests that the solar wind is powered to a large extent by an Alfv\'en-wave (AW) energy flux. AWs energize the solar wind via two mechanisms: heating and work.  We use high-resolution direct numerical simulations of reflection-driven AW turbulence (RDAWT) in a fast-solar-wind stream emanating from a coronal hole to investigate both mechanisms. In particular, we compute the fraction of the AW power at the coronal base ($P_{\rm AWb}$) that is transferred to solar-wind particles via heating between the coronal base and heliocentric distance~$r$, which we denote $\chi_{\rm H}(r)$, and the fraction that is transferred via work, which we denote~$\chi_{\rm W}(r)$. We find that $\chi_{\rm W}(r_{\rm A})$ ranges from 0.15 to~0.3, where $r_{\rm A}$ is the Alfv\'en critical point. This value is small compared to~one because the Alfv\'en speed $v_{\rm A} $ exceeds the outflow velocity~$U$ at $r<r_{\rm A}$, so the AWs race through the plasma without doing much work.  At $r>r_{\rm A}$, where $v_{\rm A} < U$, the AWs are in an approximate sense ``stuck to the plasma,'' which helps them do pressure work as the plasma expands. However, much of the AW power has dissipated by the time the AWs reach $r=r_{\rm A}$, so the total rate at which AWs do work on the plasma at $r>r_{\rm A}$ is a modest fraction of $P_{\rm AWb}$.  We find that heating is more effective than work at $r<r_{\rm A}$, with $\chi_{\rm H}(r_{\rm A})$ ranging from 0.5 to~0.7. The reason that $\chi_{\rm H} \geq 0.5$ in our simulations is that an appreciable fraction of the local AW power dissipates within each Alfv\'en-speed scale height in RDAWT, and there are a few Alfv\'en-speed scale heights between the coronal base and~$r_{\rm A}$. A given amount of heating produces more magnetic moment in regions of weaker magnetic field. Thus, paradoxically, the average proton magnetic moment increases robustly with increasing~$r$ at $r>r_{\rm A}$, even though the total rate at which AW energy is transferred to particles at $r>r_{\rm A}$ is a small fraction of~$P_{\rm AWb}$.
\end{abstract}

\vspace{0.2cm} 
\section{Introduction}
\label{sec:intro}
\vspace{0.2cm}

Following Parker's~(1958) \nocite{parker58} prediction that the Sun emits a supersonic wind, a number of studies attempted to model the solar wind as a spherically symmetric outflow powered by the outward conduction of heat from a hot coronal base~\citep[e.g.][]{parker65,hartle68,durney72,roberts72}. Although these studies obtained supersonic wind solutions, they were unable to reproduce the large outflow velocities measured in the fast solar wind near Earth ($700 - 800 \mbox{ km} \mbox{ s}^{-1}$), and they did not explain the origin of the high coronal temperatures ($\sim 10^6 \mbox{ K}$) upon which the models were based.

These shortcomings led a number of authors to conjecture that the solar wind is powered to a large extent by an energy flux carried by waves. Several observations support this idea, including in situ measurements of large-amplitude, outward-propagating Alfv\'en waves (AWs) in the interplanetary medium
\citep[e.g.,][]{belcher71,tumarsch95,bruno13} and remote observations of AW-like motions in the low corona that carry an energy flux sufficient to power the solar wind \citep{depontieu07}.

These observations have stimulated numerous theoretical investigations of how AWs might heat and accelerate the solar wind \citep[see, e.g.,][and references therein]{hansteen_velli12}. In many of these models, a substantial fraction of the Sun's AW energy flux is transferred to solar-wind particles by some form of
dissipation. Because photospheric motions primarily launch large-wavelength AWs, and because large-wavelength AWs are virtually dissipationless, the AWs are unable to transfer their energy to the plasma near the Sun unless they become turbulent. Turbulence dramatically enhances the rate of AW dissipation because it causes AW energy to cascade from large wavelengths to small wavelengths where dissipation is rapid. One of the dominant nonlinearities that gives rise to AW turbulence is the interaction between counter-propgating AWs \citep{iroshnikov63, kraichnan65}.  Because the Sun launches only outward-propagating waves, solar-wind models that invoke this nonlinearity require some source of inward-propagating AWs. One such source is AW reflection arising from the radial variation in the Alfv\'en speed \citep{heinemann80,velli93,hollweg07}. Direct numerical simulations of reflection-driven AW turbulence (RDAWT) in a 
fast-solar-wind stream emanating from a coronal  hole~\citep{perez13,vanballegooijen16,vanballegooijen17,chandran19} have shown that AW turbulence initiated by wave reflections can drive a vigorous turbulent cascade. The turbulent dissipation rates in the simulations of \cite{perez13} and~\cite{chandran19} are consistent with the turbulent heating rates in solar-wind models that rely on RDAWT, which have proven quite successful at explaining solar-wind observations
\citep{cranmer07,verdini10,chandran11,vanderholst14,usmanov14}. \citep[We note that there are a number of alternative approaches to incorporating AWs into solar-wind models --- see, e.g.,][]{suzuki06,suzuki06b, ofman10, shoda19}.

In addition to solar-wind heating via the cascade and dissipation of AW energy, AWs energize the solar wind through the work done by the AW pressure force, which directly accelerates plasma away from the Sun \citep[e.g.,][]{hollweg73b,wang93}. The relative importance of AW heating and AW work, however, is not well understood. Our goal in this paper is to use direct numerical simulations of~RDAWT to determine the fraction of the AW power that is transferred to the solar wind via turbulent dissipation between the coronal base and radius~$r$, denoted $\chi_{\rm H}(r)$, and the fraction that is transferred via work,~$\chi_{\rm W}(r)$. \footnote{It is worth noting that the thermal energy generated by AW dissipation is later converted to bulk-flow kinetic energy as the solar wind expands. That process, however, does not concern us here. Our focus is to compare the work done by AWs with the heating that results from AW dissipation.}

In Section~\ref{sec:transfer}, we present the mathematical framework that we use to address this problem and derive mathematical expressions for $\chi_{\rm H}(r)$ and $\chi_{\rm W}(r)$.  In Section~\ref{sec:sim} we compute $\chi_{\rm H}(r)$ and $\chi_{\rm W}(r)$ using high-resolution numerical simulations
of RDAWT with fixed, observationally constrained, model radial profiles for the plasma density, solar-wind outflow velocity, and background magnetic-field strength. We compare our numerical results with the values of~$\chi_{\rm H}(r)$ and $\chi_{\rm W}(r)$ that result from a previously published analytic model of RDAWT~\citep{chandran09c}. We also investigate the radial evolution of the average proton magnetic moment $k_{\rm B} T_{\perp \rm p}(r)/B(r)$ when RDAWT is the dominant proton heating mechanism, where $T_{\perp \rm p}$ is the perpendicular proton temperature, $k_{\rm B}$ is the Boltzmann constant, and $B$ is the magnetic-field strength.

\section{Energization of the solar wind by AW turbulence}
\label{sec:transfer}

We consider turbulent fluctuations within a narrow magnetic flux tube of cross-sectional area~$a(r) \ll r^2$, where~$r$ is heliocentric distance. We neglect solar rotation and take this flux tube to be centered on a radial magnetic-field line. We take the density~$\rho$, solar-wind outflow velocity~$\bm{U}$, and background magnetic field~$\bm{B}_0$ to be fixed functions of~$r$,
\begin{eqnarray} 
  \rho = \rho(r),\quad\bm{U} &=& U(r)\eb,\quad \bm{B}_0=B_0(r)\eb,
  \label{eq:radial}
\end{eqnarray}
where $\eb\equiv\vec B_0/B_0$. Magnetic-flux conservation implies that
\begin{equation}
  a(r) = \frac{B_\odot}{B_0(r)}a_\odot\label{eq:bexpansion},
\end{equation}
where $a_\odot$ and $B_\odot$  are the values of~$a(r)$ and~$B(r)$ at the coronal base, located at $r\simeq R_{\odot}$, where $R_\odot$ is the solar radius. We allow for super-radial expansion of the magnetic field, but assume that~$a/r^2$ is small enough that~$\eb $ is nearly radial and $r-R_\odot$ is approximately equal to distance from the coronal base measured along the magnetic field. We also assume that the fluctuations in the velocity and magnetic field, denoted~$\delta \bm{v}$ and~$\delta \bm{B}$, are orthogonal to~$\bm{B}_0$, that~$\delta \bm{v}$ is divergence-free, and that density fluctuations are negligibly small.

Given these simplifying assumptions, the fluctuations satisfy the equations~\citep[see, e.g.,][]{cranmer05,verdini07,chandran09c}
\begin{equation}
  \frac{\partial\vec{z}^\pm}{\partial t}+\left(U\pm
  v_A\right)\frac{\partial \vec{z}^\pm}{\partial r}+\left(U\mp
  v_A\right)\left(\frac 1{4H_\rho}\vec{z}^\pm-\frac
  1{2H_{A}}\vec{z}^\mp\right)
 =-\left(
   \vec{z}^\mp\cdot\vec{\nabla}\vec{z}^\pm -\frac{\nabla P}\rho\right)
   + \bm{D}^\pm,\label{eq:ch09} 
\end{equation}
where $\vec z^\pm=\delta\vec v\mp\delta\vec b$ are the Elsasser variables, $\delta\vec v$ is the fluctuating velocity, $\delta\vec b=\delta\vec B/\sqrt{4\pi\rho}$ is the fluctuating Alfv\'en velocity, $v_{\rm A} = B_0 /\sqrt{4\pi \rho}$ is the Alfv\'en speed, $P=p+B^2/8\pi$ is the combined plasma and magnetic pressure, $\bm{D}^\pm$ is a dissipation term that accounts for viscosity and resistivity (or possibly  hyper-viscosity and hyper-resistivity in some numerical models), and
\begin{equation}
   \frac 1{H_\rho} \equiv-\frac 1{\rho}\diff{\rho}{r},\quad\frac
                            1{H_B} \equiv -\frac
                            1{B_0}\diff{B_0}{r},\quad\hbox{and}\quad\frac 1{H_{A}} \equiv \frac 1{v_A}\diff{v_{\rm A}}{r}=\frac 1{2H_\rho}-\frac 1{H_B} 
   \end{equation}
are characteristic length scales of $\rho$, $B_0$ and $v_{\rm A}$ along the magnetic field, respectively. Our definition of Elsasser variables
with the ``$\mp$'' sign convention implies that $\vec z^+$ ($\vec z^-$) represents AW fluctuations propagating parallel (anti-parallel) to $\bm{B}_0$ in the local plasma frame. For concreteness, we take $\bm{B}_0$ to point radially outward from the Sun so that $\bm{ z}^+$ ($\bm{ z}^-$) represents AW fluctuations that propagate anti-sunward (sunward) in the plasma frame. Equations~\eqref{eq:ch09} can be also seen as an inhomogeneous version of the reduced magnetohydrodynamics (RMHD) model~\citep{kadomtsev74,strauss76} with an important new piece of physics, namely the linear coupling between~$\vec z^+$ and $\vec z^-$ fluctuations that is responsible for the non-WKB\footnote{WKB stands for the well known Wentzel–Kramers–Brillouin approximation for finding solutions of linear wave equations with variable coefficients.} reflection of AWs resulting from the spatial variation of~$v_{\rm A}$.  

\subsection{Solar wind energization: how the AW power decreases with~$r$}
\label{sec:LAW} 

The background flow and turbulent fluctuations satisfy a total-energy conservation relation, which, in steady state, takes the form \citep{chandran15b}
\begin{equation}
    \div \left[(F_{\rm flow} + F_{\rm turb})\bm{\hat{b}}\right] = 0,
\label{eq:etotal}
\end{equation}
where
\begin{equation}
F_{\rm flow} = \frac{\rho U^3}{2}   + \frac{\gamma Up}{\gamma-1} - \frac{ G M_{\odot}\rho U}{r} + q_r
\end{equation}
is the enthalpy flux of the background plasma,
$G$ is the universal gravitational constant, $M_{\odot}$ is the mass of the sun, $\gamma$ is the ratio of specific heats, $q_r$ is the radial component of the heat flux,
\begin{equation}
    F_{\rm turb}(r) = F^+(r) + F^-(r) + U(r)p_{\rm w}(r)
    \label{eq:Fturb}
\end{equation}
is the enthalpy flux of the fluctuations,
\begin{equation}
    F^\pm(r)\equiv\left[U(r)\pm v_{\rm A}(r)\right]\mathcal
                   E^\pm(r)
\end{equation}
is the energy flux of the Elsasser field $\vec z^\pm$, 
\begin{equation}
    \mathcal E^\pm(r)\equiv\aave{\frac 14\rho|\vec z^\pm|^2}=\frac 14\rho(r)[z_{\rm rms}^\pm(r)]^2\label{eq:Epm}
\end{equation}
is the average energy density of Elsasser field $\vec z^\pm$, $\aave{\cdots}$ denotes a statistical ensemble average\footnote{Note that because we assume the system is statistically stationary and homogeneous at each heliocentric distance, ensemble-averaged quantities are only a function of $r$.}, rms denotes the root-mean-square (rms) value,
\begin{equation}
    p_{\rm w}(r) = \frac 12\left[{\mathcal E}^+(r) + {\mathcal E}^-(r) -{\cal E}_{\rm
  R}(r)\right]=\frac{\delta B_{\rm rms}^2(r)}{8\pi}\label{eq:pw}
\end{equation} 
is the magnetic pressure of the fluctuations, and
\begin{equation}
    \mathcal E_{\rm R}(r)\equiv\aave{\frac 12\rho  \bm{z}^+ \cdot\bm{z}^-}=\frac 12\rho(r)\left[\delta v_{\rm
                                 rms}^2(r)-\frac{\delta B^2_{\rm rms}(r)}{4\pi\rho(r)}\right]\label{eq:ER}
\end{equation}
is the average residual energy density.

Upon taking the dot product of equation~\eqref{eq:ch09} with
$\rho\vec z^\pm/2$, performing an ensemble average, and summing the $+$ and~$-$ versions of the equation, we obtain a separate equation describing the evolution of the fluctuation energy in steady state,
\begin{equation}
 \div[(F^+ + F^-)\bm{\hat{b}}] =- p_{\rm w}\div(\bm{\hat{b}}U)-\sigma U\mathcal E_{\rm R}- Q,\label{eq:AWenergy}
 \end{equation}
 where
 \begin{equation}
   Q=\aave{\frac 12\rho(\vec z^+\cdot\vec D^++\vec z^-\cdot\vec D^-)}\label{eq:Qave}
 \end{equation}
 is the average turbulent heating rate, and $\sigma=\nabla\cdot\vec{\hat b}=1/H_B$. When ${\mathcal E}_{\rm R} \rightarrow 0$, ${\mathcal E}^-/{\mathcal E}^+ \rightarrow 0$, and $Q\rightarrow 0$, \eqref{eq:AWenergy} reduces to Equation~(42) of~\cite{dewar70}, which is equivalent to the statement that the action of outward-propagating AWs is conserved. Combining \eqref{eq:Fturb} and~\eqref{eq:AWenergy}, we obtain the relation
\begin{equation}
    \div(F_{\rm turb}\bm{\hat{b}}) = \frac{1}{a(r)}\diff{P_{\rm AW}}{r} = - \sigma U {\mathcal E}_{\rm R} + U \diff{p_{\rm w}}{r} - Q,
    \label{eq:Dewar70}
\end{equation}
where
\begin{equation}
    P_{\rm AW}(r) = a(r) F_{\rm turb}(r)
    \label{eq:defPAW}
\end{equation}
is the AW power. For the solar case, in which~${\mathcal E}^+ > {\mathcal E}^-$ and $P_{\rm AW} > 0$, \eqref{eq:Dewar70} describes how the AW power decreases with increasing~$r$. Equation~\eqref{eq:etotal} implies that $P_{\rm AW}(r) + P_{\rm flow}(r)$ is independent of~$r$, where $P_{\rm flow} = a(r) F_{\rm flow}(r)$ is the power carried by the outflowing background plasma. Thus, as $P_{\rm AW}$ decreases, $P_{\rm flow}$ increases, and energy is transferred from the AW fluctuations to the background without loss.

By multiplying~\eqref{eq:Dewar70} by~$a(r)$ and integrating from the coronal base at $r=r_{\rm b}$ out to radius~$r$ we obtain
\begin{equation}
P_{\rm AW}(r)-P_{\rm AWb} = \int_{r_{\rm b}}^ra(r')U(r')\left[ \diff{}{r'}p_{\rm
    w}(r')-\sigma(r')\mathcal E_{\rm R}(r')\right]\d r'-\int_{r_{\rm b}}^r a(r')Q(r')\d r',
\label{eq:dLAW}
\end{equation}
where the ``b'' subscript indicates that the subscripted quantity ($P_{\rm AW}$ in this case) is evaluated at $r=r_{\rm b}$. The first term inside the first integral of Equation \eqref{eq:dLAW} is the negative of the radial component of the pressure force on a plasma parcel of thickness $dr$, $-a(\d p_{\rm w}/ \d r)dr$, multiplied by the radial velocity~$U$, which has the familiar force-times-velocity form of mechanical power and represents the rate at which~$P_{\rm AW}$ decreases with increasing radius due to the work done by AWs on the flow. When residual energy $\mathcal E_{\rm R}$ is negative, which has been shown to result from the nonlinear interaction of counter-propagating AWs~\citep{muller05,boldyrev11a,wang11b}, the second term in the first integral results in a small reduction of the net work done by the AW pressure. However, positive residual energy is possible in RDAWT when linear terms responsible for non-WKB reflection dominate nonlinear terms and when $\d v_{\rm A}/\d r > 0$, in which case the residual energy will be responsible for a slight increase in the net work done by the AW pressure. The second integral in Equation~\eqref{eq:dLAW} represents the rate of decrease in~$P_{\rm AW}$ due to dissipation and turbulent heating. We  rewrite Equation~\eqref{eq:dLAW} as
\begin{equation}
P_{\rm AW}(r) - P_{\rm AWb} =  - H(r) - W(r),
\label{eq:diffLAW}  
\end{equation} 
where
\begin{eqnarray}
H(r) &=& \int_{\rm r_{\rm b}}^r a(r')Q(r') \d r'
\label{eq:defH}\\
\quad W(r) &=& - \int_{\rm
  r_{\rm b}}^r a(r')U(r')\left[\diff{}{r'}p_{\rm w}(r')-\sigma(r')\mathcal E_{\rm R}(r')\right] \d r'
\label{eq:defW} 
\end{eqnarray} 
are the rates at which energy is transferred from the fluctuations to the background flow in the radial interval~$(r_{\rm b}, r)$ via heat and work, respectively. We then divide \eqref{eq:diffLAW} by $-P_{\rm AWb}$ and rearrange terms to obtain
\begin{equation}
1 = \frac{P_{\rm AW}}{P_{\rm AWb}} + \chi_{\rm H} + \chi_{\rm W},
\label{eq:chiHW} 
\end{equation} 
where 
\begin{equation}
\chi_{\rm H}(r) = \frac{H(r)}{P_{\rm AWb}} \qquad \chi_{\rm W}(r) =
\frac{W(r)}{P_{\rm AWb}}.
\label{eq:defchiHW} 
\end{equation} 
The first term on the right-hand side of \eqref{eq:chiHW} is the fraction of the coronal-base AW power~$P_{\rm AWb}$ that survives to radius~$r$. The
second term~$\chi_{\rm H}(r)$ is the fraction of $P_{\rm AWb}$ that is transferred to solar-wind particles via dissipation and heating between~$r_{\rm b}$ and~$r$. The third term~$\chi_{\rm W}$ is the fraction of $P_{\rm AWb}$ that is transferred to solar-wind particles via work in the radial interval~$(r_{\rm b}, r)$.

\section{Heat and Work fractions of AW power transferred by RDAWT}\label{sec:sim}

In this section, we compute the rates at which AW fluctuation energy is transferred to the background solar wind via heating and work in direct numerical simulations of reflection-driven AW turbulence. 
We also compute these same rates using an approximate analytic model of reflection-driven AW turbulence.  
The numerical simulations were carried out using the pseudo-spectral
Chebyshev-Fourier REFLECT code~\citep{perez13}, which solves \eqref{eq:ch09} in the narrow-flux-tube gemoetry described in the previous section.

\subsection{Numerical simulations}

We consider the three simulations labeled
Run 1, Run 2 and Run 3 in the work of~\cite{chandran19}, in which
\begin{equation}
    \rho(r) = m_{\rm p} n(r),
    \label{eq:rhoprofile}
\end{equation}
\begin{equation}
n(r) = \left(3.23 \times 10^8 x^{-15.6} + 2.51 \times
    10^6x^{-3.76} + 1.85 \times 10^5 x^{-2}\right)  \mbox{ cm}^{-3},
\label{eq:n} 
\end{equation} 
\begin{equation}
B_0(r) = \left[1.5(f_{\rm max} - 1)x^{-6} + 1.5 x^{-2}\right] \mbox{ G},
\label{eq:B0} 
\end{equation} 
\begin{equation}
U(r) = 9.25 \times 10^{12} \;\frac{B_{\rm G}}{\tilde{n}} \;\mbox{cm}\;\mbox{s}^{-1} ,
\label{eq:defU} 
\end{equation} 
where $n$ is the proton number density, $m_{\rm p}$ is the proton mass, $x = r/R_\odot$,
$f_{\rm max}$ is the super-radial expansion factor
\citep{kopp76}, which we set equal to~9,  $B_{\rm G}$ is $B_0(r)$ in Gauss, and $\tilde{n}$ is $n(r)$ in
units of $\mbox{cm}^{-3}$. In all three runs, the simulation domain consists of a narrow  magnetic flux tube with a square cross section of area~$L_\perp(r)^2$ extending from the photosphere ($r=R_\odot$) out to approximately $r=21 R_\sun$. The Alfv\'en critical point in these simulations, defined as the heliocentric radius at which the local Alfv\'en speed equals the solar wind speed, is located  at $r=r_{\rm A}\simeq 11.1R_\sun$ (for a list of relevant heliocentric radii see Table~\ref{tab:t1}). Equation~\eqref{eq:bexpansion} implies that
\begin{equation}
  L_\perp(r) =L_{\perp,\rm b}\sqrt{\frac{a(r)}{a_\sun}}=L_{\perp,\rm b}
  \sqrt{\frac{B_\sun}{B(r)}}, 
\end{equation}
where $L_{\perp,\rm b}$ is the width of the simulation domain at the coronal base.

AWs are injected into the solar corona by imposing a broad spectrum of $z^+$ fluctuations at the coronal base, located at $r=1.0026R_\odot$ in our simulations (approximately 1800~km above the photosphere), which then propagate outwards and generate reflected (inward-propagating) waves that drive the turbulent cascade. A strong-turbulence spectrum at the coronal base is achieved by adding a model chromosphere just below the coronal base, with a sharp transition region modeled as a discontinuity of the background profiles. Although the model chromosphere ignores important effects, such as compressibility, it allows us to generate a strong turbulence spectrum of fluctuations driven by reflections from strong inhomogeneities in the chromosphere and the sharp transition region. 

The three simulations, which are described in greater detail in~\cite{chandran19} and a subsequent publication, differ only in the properties of the photospheric velocity field imposed at the inner boundary, namely, the root-mean-square amplitudes of the velocity fluctuations, the correlation lengths, and the correlation times. These parameters, shown in Table~\ref{tab:sims}, are chosen to investigate how the turbulence properties at each $r$ depend on the properties of the fluctuations at the coronal base within observational constraints. For instance, the width of the simulation domain at the coronal base, $L_{\perp,\rm b}$, is chosen to allow for characteristic correlation lengths of AWs launched at the base to be consistent with observational estimates between $10^3$~km to $10^4$~km~\citep{cranmer07,hollweg10,vanballegooijen16,vanballegooijen17,dmitruk02,verdini07,verdini12}.  Similarly, photospheric velocity driving with amplitude $\delta v_{\rm rms}\simeq 1.3$~km/s and characteristic times between 3~min to 10~min leads to $\delta v_{\rm rms, b}$ values  between $20$~km/s and $30$~km/s with correlation times between $0.3$~min and $1.6$~min. 
\begin{table}
  \centering
  \begin{tabular}{lccccccc}
    Parameter            &              & & Run 1   & & Run 2   & & Run 3 \\
    $\tau^+_{\rm c,p}$     &\dotline{1in} & & 3.3~min & & 9.6~min & & 9.3~min\\
    $\tau^+_{\rm c,b}$     &\dotline{1in} & & 0.3~min & & 0.3~min & & 1.6~min\\
    $L_{\perp,\rm b}$       &\dotline{1in} & & 4100~km & & 4100~km & & 16000~km \\
    $z^+_{\rms,\rm b}$      &\dotline{1in} & & 60~km/s & & 55~km/s & & 40~km/s\\
    $\delta v_{\rms,\rm b}$ &\dotline{1in} & & 31~km/s & & 28~km/s & & 21~km/s
  \end{tabular}
  \caption{Relevant simulation parameters. At the photosphere: $\tau^+_{\rm c,p}$ is the correlation time of velocity fluctuations imposed at $r=R_\odot$. At the coronal base: $\tau^+_{\rm c,p}$ is the correlation time of outward-propagating AWs ($z^+$), $L_{\perp,\rm b}$ is the perpendicular box length, $z^+_{\rms}$ is the \rms~amplitude of AWs injected at the base, and $v_{\rm rms,b}$ is the bulk velocity
  \rms. }\label{tab:sims}
\end{table}

\subsection{Analytic model of reflection-driven AW  turbulence}
\label{sec:CH09} 

\cite{chandran09c} (hereafter CH09) developed an analytic model of reflection-driven AW turbulence based on~\eqref{eq:ch09} but with the additional assumption that 
\begin{equation}
z^+_{\rm rms} \gg z^-_{\rm rms},
\label{eq:hch} 
\end{equation} 
which implies that 
\begin{equation} 
 \mathcal E^++\mathcal E^-\simeq\frac{1}{4}\rho (z^+_{\rm rms})^2.
\label{eq:Ew} 
\end{equation} 
Given~\eqref{eq:hch}, CH09 estimated the turbulent heating rate to be
\begin{equation}
Q = \frac{\rho z^-_{\rm rms} (z^+_{\rm rms})^2}{4 \lambda_\perp},
\label{eq:Q} 
\end{equation} 
where
$\lambda_\perp$ is the correlation length of the turbulence measured
perpendicular to~$\bm{B}_0$. Following \cite{dmitruk02}, 
CH09 estimated $z^-_{\rm rms}$ by
balancing the rate at which $\bm{z}^-$ is produced by non-WKB reflection
against the rate at which $\bm{z}^-$ cascades to small scales and
dissipates in the small-$\lambda_\perp$ limit, obtaining
\begin{equation}
z^-_{\rm rms}(r) = \frac{\lambda_\perp (U+v_{\rm A})}{v_{\rm A}}\left|
  \diff{v_{\rm A}}{r}\right|,
\label{eq:zminus} 
\end{equation} 
which is independent of~$z^+_{\rm rms}$, because the source and sink
terms for~$\bm{z}^-$ are both proportional to $z^+_{\rm rms}$.
CH09 then used~\eqref{eq:hch} and~\eqref{eq:zminus} to solve~\eqref{eq:Dewar70},
obtaining
\begin{equation}
\left[z^+_{\rm rms}(r)\right]^2 =(z^+_{\rm rms,b})^2\,\frac{\eta^{1/2}}{\eta_{\rm b}^{1/2}} \left(\frac{1+\eta_{\rm b}^{1/2}}{1+\eta^{1/2}}\right)^2
h(r),
\label{eq:zplusCH09} 
\end{equation}
where 
\begin{equation}
\eta(r) = \frac{\rho(r)}{\rho(r_{\rm A})},
\label{eq:defeta} 
\end{equation} 
\begin{equation}
h(r)=\left\{ \begin{array}{cc}
v_{\rm Ab}/v_{\rm A}(r) & \mbox{ \hspace{0.3cm} if $r_{\rm b} < r<
                        r_{\rm m}$} \vspace{0.2cm} \\ 
v_{\rm Ab}v_{\rm A}(r)/v_{\rm Am}^2 &\mbox{ \hspace{0.3cm} if $r > r_{\rm
                                   m}$} 
\end{array} \right. ,
\label{eq:defg} 
\end{equation} 
and $v_{\rm Am}$ is the maximum Alfv\'en speed in the corona, which occurs at $r=r_{\rm m}$. (CH09 assumed that $v_{\rm A}$ increased monotonically with increasing~$r$ between $r_{\rm b}$ and $r_{\rm m}$, and then decreased monotonically with increasing~$r$ at $r>r_{\rm m}$.) The CH09 model reduces to the model of \cite{dmitruk02} in the limit~$U\rightarrow 0$.

\begin{table}
\begin{center}
\begin{tabular}{cccc}
Quantity & & Meaning & Numerical value in CH09\\
& & & \\
$r_{\rm b}$ &\dotline{1in} & radius of coronal base & $1.0026 R_{\rm s}$ \\
$r_{\rm m}$ &\dotline{1in} & radius of Alfv\'en-speed maximum & $1.72 R_{\rm s}$\\
$r_{\rm A}$ &\dotline{1in} & radius of Alfv\'en critical point & $11.1 R_{\rm s}$ \\
$v_{\rm Ab}$ &\dotline{1in} & Alfv\'en speed at $r=r_{\rm b}$ & $906 \mbox{ km}
\mbox{ s}^{-1}$ \\
$v_{\rm Am}$ &\dotline{1in} & Alfv\'en speed at $r=r_{\rm m}$ & $2430 \mbox{ km}
\mbox{ s}^{-1}$ \\
$v_{\rm Aa}$ &\dotline{1in} & Alfv\'en speed at $r=r_{\rm A}$ & $626 \mbox{ km} \mbox{ s}^{-1}$ \\
$U(1 \mbox{ au})$ &\dotline{1in} & solar-wind outflow
                                                  velocity at Earth & $750
                                                                 \mbox{
                                                                 km}
                                                                 \mbox{
                                                                 s}^{-1}$
\end{tabular}
\caption{Glossary of relevant quantities and their numerical values.\label{tab:t1} }
\end{center}
\end{table}

Mass and flux conservation imply that
\begin{equation}
    \frac{\rho U}{B_0} = \mbox{constant}.
    \label{eq:massflux}
\end{equation}
This equation and the
condition that $U(r_{\rm A}) = v_{\rm A}(r_{\rm A})$ imply that
\begin{equation}
v_{\rm A} = \eta^{1/2} U.
\label{eq:vAU} 
\end{equation} 
The density at the coronal base exceeds the density at the Alfv\'en
critical point by a large factor ($\simeq 10^5$ in the fast-solar-wind
model of \cite{chandran11}). Thus,
\begin{equation}
1+\eta_{\rm b}^{1/2} \simeq \eta_{\rm b}^{1/2}.
\label{eq:etab} 
\end{equation} 
Given \eqref{eq:defPAW}, \eqref{eq:hch},~\eqref{eq:vAU}, and~\eqref{eq:etab},
\begin{equation}
P_{\rm AWb}  \simeq 
\frac{1}{4} a_{\rm b} v_{\rm Ab} \rho_{\rm b} (z^+_{\rm rms,b})^2 .
\label{eq:LAWb} 
\end{equation} 
Upon substituting \eqref{eq:Q}, \eqref{eq:zminus},
\eqref{eq:zplusCH09}, and \eqref{eq:LAWb} into \eqref{eq:defH} and
\eqref{eq:defchiHW}, we obtain
\begin{equation}
\chi_{\rm H}(r) = \int_{r_{\rm b}}^r
\left[\frac{\eta(r')^{1/2}}{1+\eta(r')^{1/2}}\right] \frac{1}{v_{\rm A}(r')}
\left| \diff{v_{\rm A}(r')}{r'}\right| h(r') \d r'.
\label{eq:chi_H_int} 
\end{equation}
Equations \eqref{eq:defPAW}, \eqref{eq:zplusCH09}, and \eqref{eq:etab} imply that
\begin{equation}
\frac{P_{\rm AW}}{P_{\rm AWb}} = \left(\displaystyle \frac{3}{2} +
  \eta^{1/2}\right)\eta^{1/2}(1 + \eta^{1/2})^{-2} h(r).
\label{eq:LAW_an} 
\end{equation} 
It then follows from \eqref{eq:chiHW} and \eqref{eq:LAW_an} that
\begin{equation}
\chi_{\rm W} = 1 - \chi_{\rm H} - \left(\displaystyle \frac{3}{2} +
  \eta^{1/2}\right)\eta^{1/2}(1 + \eta^{1/2})^{-2} h(r).
\label{eq:chiW_val} 
\end{equation}

\begin{figure}
\includegraphics{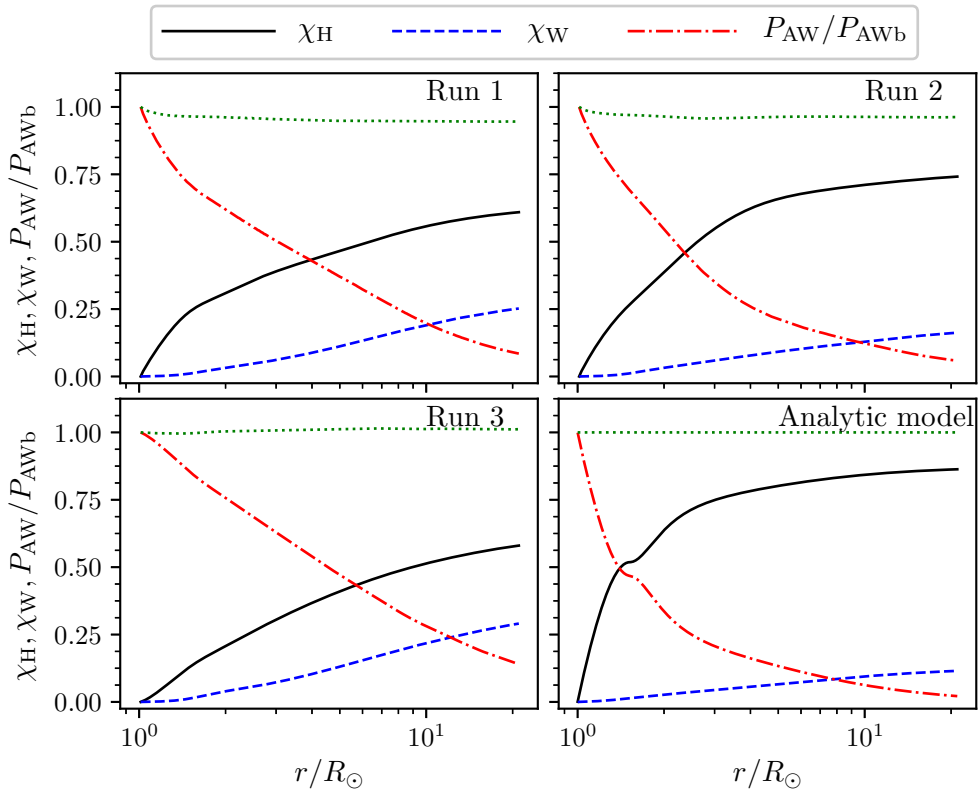}
\caption{$\chi_{\rm H}(r)$ (solid) and~$\chi_{\rm W}(r)$ (dashed) are the fractions of the Sun's AW power injected at the base that are transferred to solar-wind particles via heating and work, respectively, between the coronal base and heliocentric distance~$r$. $P_{\rm AW}/P_{\rm AWb}$ (dashed-dotted) is the fraction of the power that remains at each heliocentric distance $r$. These fractions are evaluated for Run 1, Run 2, and Run 3 using the expressions in Section~\ref{sec:LAW}
and for the CH09 analytic model (lower-right panel) using the expressions in Section~\ref{sec:CH09}. All four panels are computed using the $n(r)$, $B_0(r)$, and $U(r)$ profiles in \eqref{eq:n} through~\eqref{eq:defU}. The green dotted lines represent the sum of the three fractions, which due to energy conservation equal one in steady state. Small deviations from one in the numerical simulations are due primarily to averaging over a finite number of realizations rather than a full ensemble representing a true statistical state.
\label{fig:XHXW} }
\end{figure}

\subsection{The fractions of the AW power that are converted into heat and work}
\label{sec:fractions} 

Figure~\ref{fig:XHXW} shows the fractions $\chi_{\rm H},\chi_{\rm W}$ and $P_{\rm AW}/P_{\rm AWb}$ for Run 1, Run 2 and Run 3, as well as the corresponding fractions calculated from the analytic model described in the previous section, using equations~\eqref{eq:n},~\eqref{eq:B0},~\eqref{eq:defU},~\eqref{eq:chi_H_int},~\eqref{eq:LAW_an},~\eqref{eq:chiW_val}, and the numerical values of relevant parameters listed in table~\ref{tab:t1}. In the simulations, the heating fractions are calculated using equations~\eqref{eq:Epm}, \eqref{eq:pw}, \eqref{eq:ER}, \eqref{eq:Qave}, \eqref{eq:defH} and \eqref{eq:defW}. Assuming ergodicity in space and time at each radius, ensemble averages in equations~\eqref{eq:Epm}, \eqref{eq:ER} and~\eqref{eq:Qave} are computed using a combined average over time (during the simulation's steady state) and over the cross-sectional area of the fluxtube. More precisely, ensemble averages of any quantity $f=f(\vec x,t)$ in the simulations are computed as
\begin{equation}
  \aave{f}\equiv\frac 1T\int_0^T\left(\frac 1{a}\int_Sfda\right) dt,
\end{equation}
where $S$ is the cross-sectional surface of the fluxtube at each radius.

The reason that $\chi_{\rm H}(r_{\rm A})$ is greater than 0.5 in the simulations is that a moderate fraction of the remaining AW energy flux dissipates each time the outward-propagating AWs pass through one Alfv\'en-speed scale height \citep{dmitruk02, chandran09c}, and there are a few Alfv\'en-speed scale heights between $r=r_{\rm b}$ and~$r=r_{\rm A}$.  On the other hand, the work fraction $\chi_{\rm W}\lesssim0.3$ is significantly smaller because $v_{\rm A} > U$ at $r<r_{\rm A}$, and thus AWs at $r<r_{\rm A}$ are in an approximate sense speeding through a quasi-stationary background without doing much work. In contrast, at $r>r_{\rm A}$, $v_{\rm A} < U$, and the AWs can be thought of as being ``stuck to the plasma,'' which enhances the rate at which the fluctuations do work and causes work to become somewhat more efficient than heating. For example, $\chi_{\rm W}(21 R_{\odot}) - \chi_{\rm W}(r_{\rm A})$ equals 0.05, 0.027, 0.062, and 0.017 in Run~1, Run~2, Run~3, and the analytic model, respectively, whereas 
$\chi_{\rm H}(21 R_{\odot}) - \chi_{\rm H}(r_{\rm A})$ equals 0.04, 0.026, 0.055, and 0.016 in Run~1, Run~2, Run~3, and the analytic model, respectively. Although work is slightly more efficient than heating at transferring
energy from AWs to solar-wind particles between $r=r_{\rm A}$ and
$r=21R_{\odot}$, most of the AW energy flux has dissipated by the
time the AWs reach~$r_{\rm A}$, and the amount of AW power that is
transferred to particles via work in this region is only a tiny
fraction ($\lesssim 6\%$) of~$P_{\rm AWb}$.

The different efficiencies of AW energy loss via work inside and
outside the Alfv\'en critical point are in some ways analogous to the
different rates at which energetic particles lose energy in the
expanding solar wind in the scatter-free and scatter-dominated
regimes \citep{ruffolo95}. When pitch-angle scattering is weak, energetic particles race
through the plasma, their energies are approximately conserved, and
they do negligible work on the plasma. In contrast, when pitch-angle
scattering is strong, energetic particles are 
``stuck to the plasma,'' and they lose energy through adiabatic expansion,
because the scattering centers that ``collide'' with the particles are
rooted in the plasma and diverge from one another as the plasma
expands. As the particles lose energy, they do work on the background flow.

\begin{figure}
\begin{center}
\includegraphics{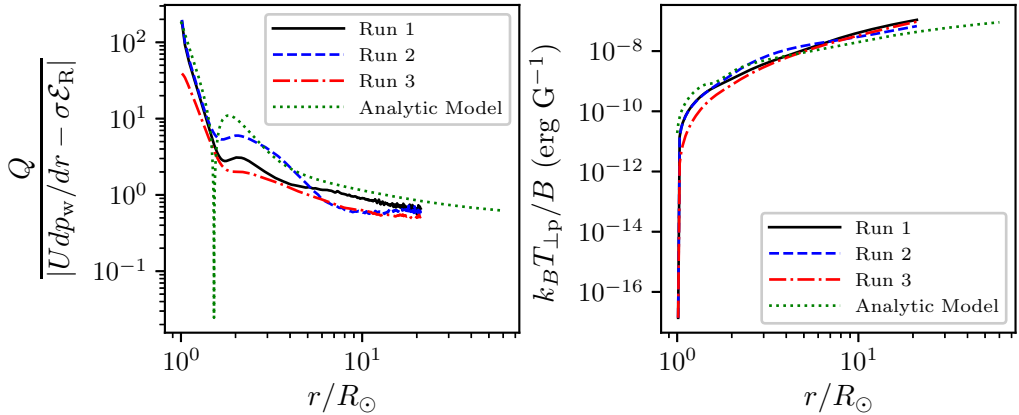}
\caption{Left panel: The ratio of the heating rate~$Q$ to the rate at which AWs do work on the flow per unit volume,~$|U \d p_{\rm w}/\d r-\sigma\mathcal E_{\rm R}|$, as a
  function of heliocentric distance~$r$. Heating is much more efficient than work close to the Sun, but at $r>r_{\rm A} = 11. 1R_{\odot}$, work becomes slightly more efficient than heating. Right panel: The average proton magnetic moment, $k_{\rm B} T_{\perp \rm p}/B$, as a function of heliocentric
  distance~$r$, computed using~\eqref{eq:Q}, \eqref{eq:zminus}, \eqref{eq:zplusCH09}, \eqref{eq:mu3}, \eqref{eq:n}, \eqref{eq:B0}, and~\eqref{eq:defU},  with $f_{\perp \rm p} =0.9$, $z^+_{\rm rms,b} = 72 \mbox{ km} \mbox{ s}^{-1}$, and $T_{\perp\rm pb} = 10^6 \mbox{ K}$.
\label{fig:PHPW} }
\end{center}
\end{figure}

In the left panel of Figure~\ref{fig:PHPW}, we plot the ratio of $Q$ to
$|U \d p_{\rm W}/\d r - \sigma U {\mathcal E}_{\rm R}|$. These two quantities
are the rates of heating and work per
unit volume, which appear on the right-hand side of \eqref{eq:dLAW}.
This figure further illustrates the increasing relative
efficiency of work beyond the Alfv\'en critical point, where the AWs
become less mobile.  The sharp feature in 
$Q/|U \d p_{\rm W}/\d r - \sigma U {\mathcal E}_{\rm R}|$ at $r=r_{\rm m} = 1.72 R_{\odot}$ in the analytic model (in which ${\cal E}_{\rm R}$ is taken to be negligible in comparison to~$p_{\rm W}$) is an artifact of the local nature of the CH09
model, in which $z^-_{\rm rms}$ is determined at each~$r$ by balancing
the local rate of wave reflection against the local rate at which
$\bm{z}^-$ fluctuations cascade and dissipate. This local balance
causes $z^-_{\rm rms}$ and~$Q$ to be proportional to $|\d v_{\rm A}/\d r|$, which vanishes at
  $r=r_{\rm m}$. In  our numerical simulations, $\bm{z}^-$ fluctuations
  propagate some radial distance before dissipating, and
  $z^-_{\rm rms}$ and~$Q$ remain nonzero at~$r=r_{\rm m}$, as seen in the $Q$ profiles for numerical simulations shown in figure~\ref{fig:PHPW}.

\subsection{Magnetic-moment production}
\label{sec:mu} 

Beyond the sonic point at $r=r_{\rm s}$ ($r_{\rm s} \simeq 2 R_{\odot}$ in coronal holes), the solar-wind outflow velocity
exceeds the proton thermal speed~$v_{\rm Tp}$, and the expansion time
scale $r/U$ becomes shorter than the minimum time in which heat can
conduct over a distance~$r$, which is~$\sim r/v_{\rm Tp}$. Proton
thermal conduction can thus be neglected to a reasonable approximation
at $r>r_{\rm s}$. For simplicity, in this section, we neglect proton
thermal conduction at all~$r$.  We also neglect energy transfer from
proton-electron collisions, as well as temperature isotropization from
collisions and instabilities.  The rate at which the average proton
magnetic moment $k_{\rm B} T_{\perp \rm p}/B$ increases with~$r$ is
then given by \citep{sharma06a,chandran11}
\begin{equation}
B n U \diff{}{r} \left(\frac{k_{\rm B} T_{\perp \rm p}}{B}\right) =
f_{\perp \rm p} Q,
\label{eq:mu1} 
\end{equation} 
where $k_{\rm B}$ is the Boltzmann
constant, $T_{\perp \rm p}$ is the perpendicular proton temperature,
and $f_{\perp \rm p}(r)$ is the fraction of the heating rate that goes
into perpendicular proton heating at heliocentric distance~$r$. We approximate \eqref{eq:mu1} by setting
$B(r) = B_0(r)$. Upon dividing~\eqref{eq:mu1} by $B_0 nU$, integrating,
and making use of~\eqref{eq:massflux}, we obtain
\begin{equation}
\frac{k_{\rm B} T_{\perp \rm p}}{B_0} = 
\frac{k_{\rm B} T_{\perp \rm pb}}{B_{0b}} + \int_{\rm r_{\rm b}}^r
\frac{f_{\perp \rm p}(r^\prime)
  Q(r^\prime)  B_{0 \rm b}}{n_{\rm b} U_{\rm b} 
  [B_0(r^\prime)]^2} \d r^\prime.
\label{eq:mu3} 
\end{equation} 
The form of the integrand in~\eqref{eq:mu3} shows that a given amount of heating produces
more magnetic moment in regions of weaker magnetic field.

In right panel of figure~\ref{fig:PHPW}, we plot the average magnetic moment $k_{\rm B} T_{\perp\rm p}/B$ as a function of~$r$ using \eqref{eq:n},
\eqref{eq:B0}, \eqref{eq:defU}, \eqref{eq:Q}, \eqref{eq:zminus}, \eqref{eq:zplusCH09}, and~\eqref{eq:mu3}, with $f_{\perp \rm p} = 0.9$, $z^+_{\rm rms,b} = 72 \mbox{ km} \mbox{s}^{-1}$, and $T_{\perp \rm p b} = 10^6 \mbox{ K}$. Figures~\ref{fig:XHXW} and~\ref{fig:PHPW} show that, although only a small fraction of $P_{\rm AWb}$ is dissipated at $r>r_{\rm A}$, $k_{\rm B} T_{\perp\rm p}/B$ rises robustly at $r>r_{\rm A}$.

\section{Conclusion}

In this paper, we use direct numerical simulations of RDAWT to determine $\chi_{\rm H}(r)$ and $\chi_{\rm W}(r)$, the fractions of the AW power at the coronal base ($P_{\rm AWb}$) that are transferred to the solar wind via heating and  work between the coronal base ($r=r_{\rm b}$) and radius~$r$. Our simulations solve for the evolution of transverse, non-compressive fluctuations in a fixed background solar wind whose density, magnetic-field strength, and outflow-velocity profiles are chosen to emulate a fast-solar-wind stream emanating from a coronal hole.
We find that heating from the cascade and dissipation of AW fluctuations between $r_{\rm b}$ and the Alfv\'en critical point~$r_{\rm A}$ transfers between 50\% and 70\% of $P_{\rm AWb}$ to the solar wind, whereas work in this same region transfers between 15\% and 30\% of $P_{\rm AWb}$ to the solar wind. The variation in these numbers arises from the different photospheric boundary conditions imposed in our different numerical simulations.

The reason that $\chi_{\rm H}(r_{\rm A})$ is in the range of 50\% to 70\%
is that a moderate fraction of the local AW power dissipates within each Alfv\'en speed scale height \citep{dmitruk02,chandran09c}, and there are a few Alfv\'en speed scale heights between~$r_{\rm b}$ and~$r_{\rm A}$. The reason that $\chi_{\rm W}(r_{\rm A})$ is small compared to~1 is that $v_{\rm A} > U$ at $r<r_{\rm A}$, so AWs in this sub-Alfv\'enic region are in an approximate sense speeding through a quasi-stationary background without doing much work. Work becomes relatively more efficient at transferring AW energy to the particles at $r>r_{\rm A}$, where $v_{\rm A}< U$ and the AWs are in an approximate sense ``stuck to the plasma.''  However, because most of the Sun's AW power dissipates via heating before the AWs reach~$r_{\rm A}$, the total rate at which work transfers AW energy to the plasma at $r>r_{\rm A}$ is a small fraction of~$P_{\rm AWb}$. Although only a small fraction of~$P_{\rm AWb}$ survives to reach~$r_{\rm A}$, the average proton magnetic moment increases robustly at~$r>r_{\rm A}$ (assuming that a substantial fraction of the turbulent heating rate goes into perpendicular proton heating at these radii), because heating becomes more effective at producing magnetic moment in regions of weaker magnetic field.

The accuracy of our results is limited by our neglect of compressive fluctuations, which enhance the dissipation of AW energy via ``AW phase mixing'' \citep{heyvaerts83}. Plasma compressibility further enhances the rate of AW dissipation via parametric decay \citep{galeev63,sagdeev69,cohen74,goldstein78,spangler86,hollweg94,dorfman16,tenerani17,chandran18a}. Three-dimensional compressible MHD simulations of the turbulent solar wind from $r=r_{\rm b}$ out to $r> r_{\rm A}$, such as those carried out by \cite{shoda19}, could lead to improved estimates of~$\chi_{\rm H}(r)$ and $\chi_{\rm W}(r)$.

\acknowledgements
\textbf{Acknowledgements.}

We thank Joe Hollweg, Mary Lee, Brian Metzger, and Eliot Quataert for valuable
discussions. JCP was supported by NSF grant AGS-1752827, BDGC was supported in part by NASA grants NNX17AI18G and 80NSSC19K0829 and NASA grant NNN06AA01C to the Parker Solar Probe FIELDS Experiment. KGK was supported in part by NASA ECIP grant 80NSSC19K0912 and the Parker Solar Probe SWEAP contract NNN06AA01C. MM was supported in part by NASA grant 80NSSC19K1390. An award of computer time was provided by the Innovative and Novel Computational Impact on Theory and Experiment (INCITE) program. During the INCITE award period (from 2012 to 2014), this research used resources of the Argonne Leadership Computing Facility, which is a DOE Office of Science User Facility supported under Contract DE-AC02-06CH11357. This work also used high-performance computing resources of the Texas Advanced Computing Center (TACC) at the University of Texas at Austin, under project TG-ATM100031 of the Extreme Science and Engineering Discovery Environment (XSEDE), which is supported by National Science Foundation grant number ACI-1548562.

\textbf{Declaration of interests.}

The authors report no conflict of interest.


\bibliographystyle{jpp}

\end{document}